\documentclass[a4paper,12pt]{article}

\usepackage{graphicx}
\usepackage{upgreek}
\usepackage{amsmath}
\usepackage{amssymb}
\usepackage{enumerate}
\usepackage{fullpage}
\usepackage{verbatim}
\usepackage{hyperref}

\pdfoutput = 1

\hypersetup{
    bookmarks=true,         
    unicode=false,          
    pdftoolbar=true,        
    pdfmenubar=true,        
    pdffitwindow=false,     
    pdfstartview={FitH},    
    pdftitle={Quantum interference between two single photons of different microwave frequencies},    
    pdfauthor={},     
    pdfsubject={Quantum interference between two single photons of different microwave frequencies},   
    pdfcreator={},   
    colorlinks=true,       
    linkcolor=red,          
    citecolor=red,        
    filecolor=red,      
    urlcolor=red,          
    pdfpagelayout=TwoPageLeft
}

\renewcommand{\mu}{\upmu}

\begin{document}

\title{Quantum interference between two single photons of different microwave frequencies}

\author{Fran\c{c}ois Nguyen,$^{1}$ Eva Zakka-Bajjani,$^{1}$ \\ Raymond W. Simmonds,$^{1}$ and Jos\'{e} Aumentado$^{1}$}

\date{}

\maketitle

\begin{enumerate}
 \item[$^1$] {\it National Institute of Standards and Technology, 325 Broadway, Boulder CO 80305, USA}
\end{enumerate}

\begin{abstract}
We have measured quantum interference between two single microwave photons trapped in a superconducting resonator, whose frequencies are initially about 6 GHz apart. We accomplish this by use of a parametric frequency conversion process that mixes the mode currents of two cavity harmonics through a superconducting quantum interference device, and demonstrate that a two-photon entanglement operation can be performed with high fidelity.
\end{abstract}

Two identical optical photons simultaneously sent through the two input ports of a semi-trans\-parent beam splitter will exit together through the same output port. This is a consequence of a complete destructive quantum interference between the two possible paths in which the photons leave by different outputs. This effect was first experimentally observed in parametric down-conversion \cite{hong1987measurement,Rarity1990}, and more recently by use of various single-photon sources \cite{beugnon2006quantum,PhysRevLett.96.240502,charles2002indistinguishable}. It is at the heart of dual-rail linear optical quantum computing schemes \cite{knill2001scheme,o2003demonstration}, where information is encoded on different spatial modes. However, these schemes can be extended to any type of system composed of linearly-coupled bosonic modes.

Parametric frequency conversion (PFC) enables the creation of a tunable direct coupling between quantized linear modes of different frequencies via a time-varying, or a nonlinear element pumped at their frequency difference. It realizes an active  beam splitter (BS) component between these modes of different frequencies, with an experimentally tunable mixing angle \cite{tucker1969quantum,louisell1960coupled,PhysRevLett.104.230502,zakka2011quantum}. As such, it has been proposed as a way to implement interferences \cite{raymer2010interference,langford2011efficient} between photonic quantum bits (qubits) encoded in frequency modes. In optics, PFC can be induced by intensely pumping a nonlinear crystal, and has been used to transduce a quantum state of light from one wavelength to another \cite{huang1992observation}, down to the single-photon level \cite{vandevender2004high,rakher2010quantum}. Superconducting circuits are attractive systems to manipulate photons by means of parametric processes based on the microwave modulation of a Josephson element. The dynamics of the parametric coherent exchange of a single excitation between two superconducting qubits \cite{niskanen2007quantum}, or two linear resonators \cite{zakka2011quantum}, far-detuned in frequency, has been recently measured. Parametric interactions can be combined with the tools developed for circuit quantum electrodynamics (cQED) \cite{wallraff2004strong} that enable the preparation and the read out of the quantum number states stored in a resonator using a superconducting qubit \cite{sillanpää2007coherent,hofheinz2008generation}. Here, we prepare two single photons of different frequencies in two resonant modes of a superconducting cavity  that are parametrically coupled by a flux-modulated superconducting quantum interference device (SQUID). We demonstrate quantum interference between these two single photons by implementing a Mach-Zehnder interferometer based on this parametric mode-mixing. Finally, we characterize the efficiency of the photon-pair preparation by measuring the fidelity of the two-photon entangled state created after interference on a semi-transparent (50:50) BS, and also give, as a comparison, the results obtained for single-photon interference.

Our circuit consists of a $\lambda$/4 coplanar stepped-impedance resonator terminated to ground through a SQUID (Fig. 1a and 1b) \cite{zakka2011quantum}. The other end is capacitively coupled to a $50~\Omega$ transmission line to allow spectroscopic measurements of the cavity by microwave reflectometry. At this node, the cavity is also coupled with a strength $g_{\textrm{q1}}$ = 17 MHz to a flux-biased phase qubit that is used to load (and detect) microwave photon number states in (from) one mode of the cavity \cite{hofheinz2008generation}. An inductively coupled coil on-chip, referred as the ``pump'', allows the application through the SQUID loop of a global flux $\Phi_{\textrm{sq}} = \Phi_{\textrm{sq}}^{\textrm{dc}} + \Phi_{ \textrm{sq}}^{\textrm{$\mu$w}}(t)$ composed of a small microwave modulation $\Phi_{ \textrm{sq}}^{\textrm{$\mu$w}}(t)$ on top of a d.c. bias $\Phi_{\textrm{sq}}^{\textrm{dc}}$. The SQUID acts as a flux-dependent lumped-element inductor $L_{\textrm{sq}}(\Phi_{\textrm{sq}}^{\textrm{dc}})$ that slightly modifies one of the boundary conditions of the cavity \cite{Wallquist2006,palacios2008tunable,yamamoto2009flux}. The resulting flux dependence of the resonance frequencies $\nu_{0}$ and $\nu_{1}$ of the first two modes 0 and 1 is shown  in Fig. 1b. The experiments reported here are performed at a flux bias $\Phi_{\textrm{sq}}^{\textrm{A}} = 0.21\Phi_{0}$ (point ``A'' in Fig. 1b) where $\nu_{0}^{\textrm{A}} = 3.8296$ GHz and $\nu_{1}^{\textrm{A}} = 9.6087$ GHz. The extracted loaded quality factors are about 16000 for both modes, high enough to perform dynamical manipulation of photon states trapped in the cavity via PFC \cite{zakka2011quantum}. By applying a small microwave drive $\Phi_{ \textrm{sq}}^{\textrm{$\mu$w}}(t) = \delta\Phi \cos (2\pi\nu_{\textrm{p}}t + \varphi_{\textrm{p}})$ at a resonant pump frequency $\nu_{\textrm{p}} = \nu_{1}^{\textrm{A}} - \nu_{0}^{\textrm{A}}$, with a pump phase $\varphi_{\textrm{p}}$ and an amplitude $\delta\Phi$, PFC is induced between the modes 0 and 1, with a parametric coupling rate $g_{10}$ proportional to $\delta\Phi$. We apply a $\delta\Phi \sim 0.019 \textrm{ }\Phi_{0}$ flux modulation, which yields to $g_{10} = 11.3~\textrm{MHz}$. At a quantum level, PFC is described in terms of annihilation and creation operators $\hat{a}_{0, 1}$ and $\hat{a}^\dagger_{0, 1}$ of quantized microwave excitations, namely photons, of these harmonic modes, by the following interaction Hamiltonian \cite{tucker1969quantum} :
\begin{eqnarray}\label{hamiltonian}
\hat{H}_I =  h g_\textrm{10} (\hat{a}^\dagger_{\textrm{0}} \hat{a}_{\textrm{1}} e^{i \varphi_\textrm{p}} + \hat{a}_{\textrm{0}} \hat{a}^\dagger_{\textrm{1}} e^{-i \varphi_\textrm{p}})
\end{eqnarray}
in the doubly rotating frame defined by the unitary transformation $e^{2 i \pi [\nu_\textrm{0} \textrm{ }\hat{a}^\dagger_\textrm{0} \hat{a}_\textrm{0} + \nu_\textrm{1} \textrm{ } \hat{a}^\dagger_\textrm{1} \hat{a}_\textrm{1}] t}$. $\hat{H}_I$ describes a BS operation between modes 0 and 1, with an effective mixing angle $\theta_{\textrm{p}} = 4\pi g_{10} \Delta \tau_{\textrm{p}}$ that can be modulated by varying the duration $\Delta \tau_{\textrm{p}}$ of the parametric drive. We consider the evolution of the $\left| 0_{0} 1_{1}\right>$ and $\left| 1_{0} 1_{1}\right>$ states (the subscript denotes the mode number) under the unitary operation $U(\theta_{\textrm{p}}, \varphi_{\textrm{p}})$ associated with the parametric interaction $\hat{H}_I$, for $\theta_{\textrm{p}} \in [0;2\pi]$.
The $\theta_{\textrm{p}} = \pi/2$ operation is analogous to a 50:50 BS and creates the maximally entangled photon number state, a so-called NOON state \cite{boto2000quantum}, with one (N=1) or two (N=2) photons $\left| \psi_{\textrm{N}} \right> = \frac{f_{\textrm{N}}(\varphi_{\textrm{p}})}{\sqrt{2}}\textrm{ }\big[- e^{i\textrm{N}(\varphi_{\textrm{p}}+\pi/2)} \left|\textrm{N}_{0} 0_{1} \right>+\left|0_{0}\textrm{N}_{1}\right>\big]$, where $f_{\textrm{N}}(\varphi_{\textrm{p}})$ is a global phase. In the N=2 case, destructive two-photon interference leads to the disappearance of $\left|1_{0}1_{1}\right>$ in the final state, and the creation of $\left|\psi_{2}\right>$ illustrates the bosonic nature of the photons, which ideally exit together through either ``frequency output port''. Note that the preparation of NOON states between two superconducting resonators has been demonstrated in ref. \cite{wang2010deterministic}. The method used in their work is different, as it relies on transferring N times the single-excitation entanglement between two phase qubits into two distinct microwave resonators, the final state being characterized by a bipartite Wigner tomography.

We prepare the initial state of the cavity by use of a phase qubit \cite{LawEberly, hofheinz2008generation,sillanpää2007coherent} (Fig. 2a). The qubit frequency $\nu_{\textrm{q}}$ can be tuned from 7.5 GHz to 12.5 GHz by means of an external flux $\Phi_{\textrm{q}}$. The qubit state is manipulated with microwave pulses and measured with an inductively coupled d.c. SQUID. By applying a fast flux pulse to the qubit, the excited state $\left|\textrm{e}\right>$ state preferentially tunnels to a different flux state, which is indicated by a shift in the d.c. SQUID critical current. When the qubit is in the ground state, the tunneling probability $P_\textrm{t}$ is about $0.06$; this probability increases linearly with the $\left|\textrm{e}\right>$ state occupancy, up to $0.8$ when the qubit is prepared in its excited state. We excite the qubit far off resonance from mode 1 (at $\nu_{\textrm{q}}^{\textrm{off}} = 8.775$ GHz), then bring it into resonance in order to swap the qubit excitation to the cavity, which prepares $\left| 0_{0} 1_{1}\right>$. To prepare the $\left|1_{0}1_{1} \right>$ state, we transfer the photon in mode 1 to mode 0 by applying a $\theta_{\textrm{p}}= \pi$ pump pulse, then prepare and transfer another excitation from the qubit to mode 1. In practice, the relaxation of the qubit and of the cavity prevent us from preparing these initial pure Fock states. We prepare instead a mixed state described by the diagonal terms of an initial density matrix denoted $\rho^{\textrm{twin}}(\theta_\textrm{p}=0)$ ($\rho^{\textrm{single}}(\theta_\textrm{p}=0)$) when targeting the $\left|1_{0}1_{1}\right>$ ($\left|0_{0}1_{1}\right>$) initial state, and expressed in the $\left|\textrm{m}_{0}\textrm{n}_{1}\right>$, $\textrm{m}_{0}+\textrm{n}_{1}$ $\leq 2$ ($\textrm{m}_{0}+\textrm{n}_{1}$ $\leq 1$), photon Fock states basis. We analyze the state of mode 1 after the $U(\theta_{\textrm{p}}, \varphi_{\textrm{p}})$ evolution using the qubit as probe. After applying the $(\theta_{\textrm{p}},\varphi_{\textrm{p}})$ pump pulse, we bring the qubit into resonance with mode 1 for a varying interaction duration $\Delta\tau_{\textrm{q1}}$ and measure the qubit tunneling probability $P_{\textrm{t}}(\Delta\tau_{\textrm{q1}})$ (Fig. 2b and 2c). The probabilities $P_{\textrm{n}}(\theta_{\textrm{p}},\varphi_{\textrm{p}})$ to measure n photons in mode 1 are given by the weights of the components scaling as $\sqrt{\textrm{n}}g_{\textrm{q1}}$ in the single-photon oscillations between the qubit and the mode 1 that interact during $\Delta\tau_{\textrm{q1}}$ (The $P_{\textrm{n}}(\theta_{\textrm{p}},\varphi_{\textrm{p}})$ are extracted by simultaneous least-squares fits to $P_{\textrm{t}}(\Delta\tau_{\textrm{q1}})$ with a Lindblad-type equation). They are related to the probabilities $p_{\textrm{mn}}(\theta_{\textrm{p}},\varphi_{\textrm{p}}) = \left<\textrm{m}_{0}\textrm{n}_{1}\right|\rho(\theta_{\textrm{p}},\varphi_{\textrm{p}})\left|\textrm{m}_{0}\textrm{n}_{1}\right>$ to simultaneously measure m and n photons in the mode 0 and 1 following $P_{\textrm{n}}(\theta_{\textrm{p}},\varphi_{\textrm{p}}) = \sum_{\textrm{m}+\textrm{n}\leq \textrm{N}}\textrm{ }p_{\textrm{mn}}(\theta_{\textrm{p}},\varphi_{\textrm{p}})$, with N=1 (N=2) in the single-photon (twin-photon) case.
The extracted $P_{\textrm{n}}(\theta_{\textrm{p}}, 0)$ (the phase of the first pump pulse is set to 0 by convention and will not be specified in most cases) are shown in Fig. 2b and 2c. For both the single-photon and the twin-photon case, we measure $P_{1}(2\pi) \approx 0.92\textrm{ }P_{1}(0)$ and can reasonably neglect the effect of decoherence over a pump period to simplify the analysis. We thus infer the photon number probability of the mode 0 by use of a pump swap operation, and consider that $p_\textrm{N0}(\theta_\textrm{p}) = p_\textrm{0N}(\theta_\textrm{p}+\pi) $. The measured initial population $p_{01}(0)$ and the estimated initial population $p_{11}(0)$ are shown in Table 1. In particular, when starting with $\rho^{\textrm{twin}}(0)$, we observe a $\sqrt{2}g_{\textrm{q1}}$ component in the vacuum Rabi oscillations after a $\pi/2$ (or a $3\pi/2$) pump pulse with a weight indicating that we have prepared at least $p_{02}(\pi/2)+p_{20}(\pi/2) = 0.45 \pm 0.04$ of the $\left|1_{0} 1_{1}\right>$ state in the initial mixture.
We characterize the quality of the NOON state created after a $\pi/2$ pump pulse by performing Ramsey interferences \cite{sackett2000experimental,leibfried2004toward}(Fig. 2d). After preparing $\rho^{\textrm{single}}\big(\frac{\pi}{2}\big)$ or $\rho^{\textrm{twin}}\big(\frac{\pi}{2}\big)$, we apply a second $\pi/2$ pulse differing by a tunable phase $\Delta\varphi_{\textrm{p}}$.We measure with the qubit the final probability $P_{\textrm{n}}\big(\frac{\pi}{2}, 0; \frac{\pi}{2}, \Delta\varphi_{\textrm{p}}\big)$ to have n photons in mode 1 after this Ramsey sequence, as a function of $\Delta\varphi_{\textrm{p}}$ (Fig. 2e and 2f). The coherent superposition of $\left|0_{0}\textrm{N}_{1}\right>$ and $\left|\textrm{N}_{0}\textrm{0}_{1}\right>$ induces a sinusoidal component oscillating at N$\Delta\varphi_{\textrm{p}}$ in the Ramsey fringes pattern. More precisely, the consecutive application of $U\big(\frac{\pi}{2}, 0\big)$ and $U\big(\frac{\pi}{2}, \Delta\varphi_{\textrm{p}}\big)$ projects the coherences $c_{\textrm{0N}} = \left<0_{0}\textrm{N}_{1}|\rho\big({\frac{\pi}{2}}\big)|\textrm{N}_{0}0_{1}\right> = \big|c_{\textrm{0N}}\big|e^{i\varphi_{\textrm{0N}}}$ onto the final measured probability expressed by
\begin{eqnarray}\label{coherences}
P_{1}\bigg( \frac{\pi}{2}, 0; \frac{\pi}{2}, \Delta\varphi_{\textrm{p}}\bigg)& = &-\big|c_{01}\big|\sin(\Delta\varphi_{\textrm{p}} - \varphi_{01}) \\
&&+ \big|c_{02}\big| \cos(2\Delta\varphi_{\textrm{p}} - \varphi_{02}) + A_{1},\nonumber
\end{eqnarray}
where $A_{1}$ is a constant depending on the $p_{\textrm{mn}}\big(\frac{\pi}{2}\big)$, and with $c_{02} = 0$ in the single-photon case. The main results are summarized in Table 1. From the two types of measurements shown in Fig. 2, we can estimate the fidelity $\mathcal{F}_{\textrm{N}} = \left<\psi_{\textrm{N}}\big|\rho\big(\frac{\pi}{2}\big)\big|\psi_{\textrm{N}}\right> = \frac{1}{2}(p_{\textrm{0N}}+p_{\textrm{N0}}) + \big|c_{\textrm{0N}}\big|$ of the NOON states preparation with our protocol combining transfer of an excitation from the qubit to the cavity, and mode coupling with PFC. We find $\mathcal{F}_{\textrm{1}} = 0.74 \pm 0.04$ and $\mathcal{F}_{\textrm{2}} = 0.43 \pm 0.04$. The fidelity of the NOON state preparation is limited by the probability of initially preparing a single photon in mode 1 ($p_{01}(0)$), or a single photon in each mode ($p_{11}(0)$). The fidelity $\mathcal{F}_{\textrm{1}}$ is still higher than 0.5 by a significant amount, which is enough to claim deterministic entanglement between the two modes via a single photon \cite{sackett2000experimental}. Nevertheless, we can estimate the fidelity $\mathcal{F}_{\textrm{N}}^*$ of the N-photon entanglement operation, defined as the conditional probability $\mathcal{F}_{\textrm{N}}^*=\mathcal{F}_{\textrm{N}}/\sum_{\textrm{m}+\textrm{n} = \textrm{N}} p_\textrm{mn}(0)$, to create a NOON state, knowing that we started from one (N=1 case) or two (N=2) photons in the cavity modes (Note that in the original Hong-Ou-Mandel experiment, this type of renormalization is achieved by working in the coincidence basis \cite{hong1987measurement}). Due to relaxation of the qubit and the cavity during the pulse sequence, we have about $20~\%$ probability to lose the excitation when transferring it from the qubit into mode 1 (Table 1). This leads to $\mathcal{F}_{\textrm{1}}^* = 0.90 \pm 0.06$. In the N=2 case, we can place an upper bound on $\sum_{\textrm{m}+\textrm{n} = \textrm{2}} p_\textrm{mn}(0)$ using the probability $p_{01}^{2}(0) = 0.62 \pm 0.04$ to load sequentially two excitations from the qubit to mode 1. This implies that $\mathcal{F}_{\textrm{2}}^*$ is higher than $0.7 \pm 0.06$, which unambiguously proves the existence of two-photon quantum correlations in the created state. A more realistic upper bound of about 0.55 can be placed on $\sum_{\textrm{m}+\textrm{n} = \textrm{2}} p_\textrm{mn}(0)$ by taking into account the loss during the transfer to the mode 0 of the first photon loaded in mode 1, as well as its decay while preparing the second photon in mode 1.
\begin{table}
\caption{Characterization of single and two-photon states}
\begin{center}
\begin{tabular*}{10cm}{c c c c c}
  \hline
  &$p_{\textrm{01}}$&$p_{\textrm{10}}$&&$|c_{\textrm{01}}|$\\
  \hline
  $\rho^{\textrm{single}}(0)$&0.79(2)&0.03(2)&&-\\
  $\rho^{\textrm{single}}(\frac{\pi}{2})$&0.38(2)&0.36(2)&&0.37(2)\\
  \hline
  &$p_{\textrm{02}}$&$p_{\textrm{20}}$&$p_{\textrm{11}}$&$|c_{\textrm{02}}|$\\
  \hline
  $\rho^{\textrm{twin}}(0)$&$<0.01$&$<0.01$&$\geqslant0.45(4)^{\textrm{a}}$&-\\
  $\rho^{\textrm{twin}}(\frac{\pi}{2})$&0.22(2)&0.23(2)&&0.21(2)\\
  \hline
  \multicolumn{5}{l}{}\\
  \multicolumn{5}{l}{\parbox{10cm}{$^{\textrm{a}}$ This value is a lower boundary of $p_{\textrm{11}}(0)$ given by the sum of the $p_{\textrm{20}}(\pi/2)$ and $p_{\textrm{02}}(\pi/2)$.}}\\
  \multicolumn{5}{l}{\parbox{10cm}{$^{\textrm{b}}$ The number in parentheses is the uncertainty on the last digit.}}\\
\end{tabular*}
\end{center}
\end{table}

We also measure the decay of the NOON states by varying either the delay after a $\pi/2$ pulse in order to access the decay of the populations $p_{\textrm{0N}}\big(\frac{\pi}{2}\big)$ and $p_{\textrm{N0}}\big(\frac{\pi}{2}\big)$, or the delay between the two $\pi/2$ pump pulses of the Ramsey interferometer in order to access the decay of the coherences $c_{\textrm{0N}}$. We find that $c_{\textrm{01}}$ and $c_{\textrm{02}}$ decay exponentially with respective characteristic times $\tau_{|c_{01}|} \approx 370$ ns and $\tau_{|c_{02|}} \approx 190$ ns, close to that of the harmonic mean $2(1/\tau_{p_\textrm{N0}}+1/\tau_{p_\textrm{0N}})^{-1}$ of the relaxation times $\tau_{p_\textrm{N0}}$ and $\tau_{p_\textrm{0N}}$ of the populations (see Fig. 3), indicating that decoherence is limited by relaxation. We also observe, as expected \cite{Lu}, a decay rate of the two-photon entanglement approximately twice as fast as the single-photon one.

To conclude, we have demonstrated quantum interference between two single photons of significantly different microwave frequencies by use of PFC. The contrast of the observed interference is limited by the fidelity of the initial twin-photon state preparation. Nevertheless, the fidelity of the two-photon entanglement operation is higher than 70 $\%$. Improving the relaxation time of the qubit and of the cavity would increase the quality of the two-photon entanglement calibration, as well as the photon-pair creation efficiency. This could make this type of parametric device suitable for processing quantum information with photonic qubits encoded in microwave resonator modes.

\begin{enumerate}
\item[] {\bf Acknowledgments}
We thank N. Bergren and J. Koch for technical help. We thank D.J. Wineland for his guidance and comments. We thank T. Gerrits for his comments on the manuscript. This paper is a contribution by the National Institute of Standards and Technology and not subject to US copyright.
\item[] {\bf Author Information}
Correspondence and requests for materials should be addressed to E.Z-B. (eva.zakka-bajjani@nist.gov) or F.N. (francois.nguyen@nist.gov).
\end{enumerate}

\begin{figure}
\begin{center}
\includegraphics[scale = 0.4]{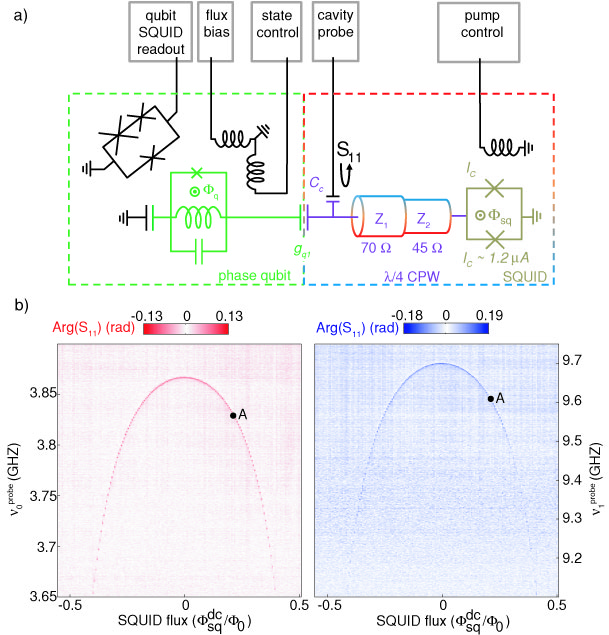}
\end{center}
\caption{(a) A $\lambda/4$ CPW resonator with a boundary condition that is varied by a SQUID is coupled to a phase qubit. The flux through the SQUID is d.c.-biased and microwave modulated via an inductively coupled pump line. The cavity is coupled to a $50\textrm{ }\Omega$ transmission line through a small capacitance $C_{\textrm{c}} \approx 1.8$ fF to allow reflectometry measurements. We manipulate the qubit state with microwave pulses and tune its frequency with a flux bias line. Measurements are performed in a dilution refrigerator operated at $35~\textrm{mK}$.
(c) Measured phase Arg$(S_{11})$ of the reflection coefficient on $C_{\textrm{c}}$ as a function of both d.c.-flux in the SQUID and probe frequency.}
\end{figure}

\begin{figure}
\begin{center}
\includegraphics[scale = 0.2]{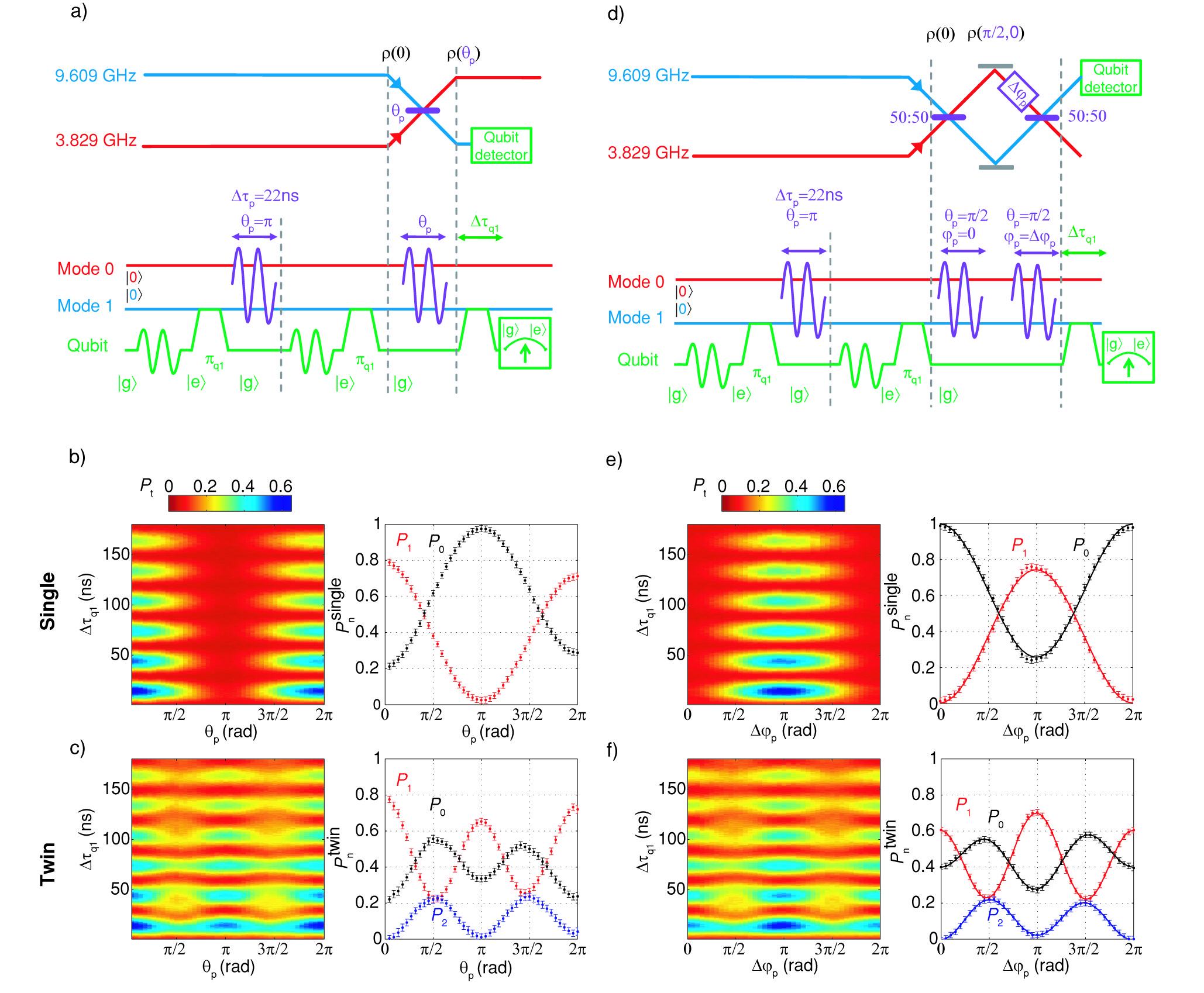}
\end{center}
\caption{(a) Microwave pulse sequence and optical analog for the states' evolution under PFC. Qubit, mode 0 and mode 1 are respectively represented in green, red and blue. The mode mixing via the parametric pump is depicted in purple.(b, c) On the left, 2D plots of the tunneling probability $P_{\textrm{t}}(\Delta\tau_{\textrm{q1}}, \theta_{\textrm{p}})$ when starting with $\rho^{\textrm{single}}(0)$ (b) or $\rho^{\textrm{twin}}(0)$ (c). On the right, the extracted probabilities $P_{\textrm{n}}$ to measure n photons in mode 1 are plotted as a function of $\theta_{\textrm{p}}$. The error bars represent both the switching measurement statistical error and standard deviations from fits to the data.
(d), Microwave pulse sequence and optical analog for the NOON states characterization. We apply a Ramsey sequence, i.e. two $\pi/2$ pump pulses differing by a phase $\Delta\varphi_{\textrm{p}}$, on the initial state.
(e, f), Ramsey fringes on the $\rho^{\textrm{single}}(0)$ (e) and on the $\rho^{\textrm{twin}}(0)$ (f) states. On the left, 2D plots of the tunneling probability $P_{\textrm{t}}(\Delta\tau_{\textrm{q1}}, \Delta\varphi_{\textrm{p}})$. On the right, the extracted probabilities $P_{\textrm{n}}$ are plotted as a function of $\Delta\varphi_{\textrm{p}}$, and fitted (solid curves) by a cosine-sine decomposition restricted to frequencies of 0, 1, ...N.}
\end{figure}

\begin{figure}
\begin{center}
\includegraphics[scale = 0.2]{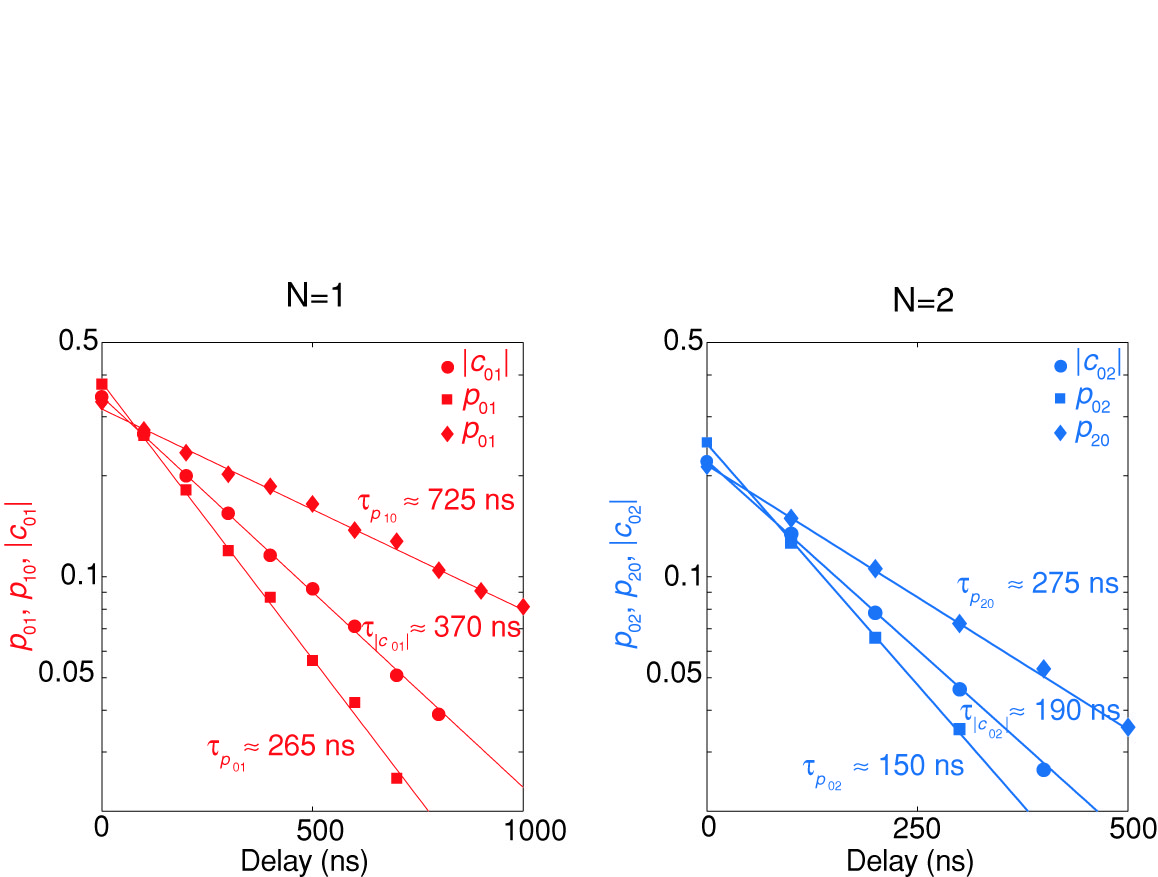}
\end{center}
\caption{Decay of the single and two-photon NOON states. The amplitudes of the coherence $c_{\textrm{0N}}$, of the probabilities $p_{\textrm{0N}}\big(\frac{\pi}{2}\big)$ and $p_{\textrm{N0}}\big(\frac{\pi}{2}\big)$, are respectively  plotted in dots, squares and diamonds, in logarithmic scale, as a function of the delay between either the two $\pi/2$ pulses of the Ramsey sequence for the coherence decay measurement, or the delay after a simple $\pi/2$ pulse for the populations decay measurements. These data are well fitted (solid curves) by an exponential decay.}
\end{figure}

\end{document}